\documentclass[twocolumn,superscriptaddress,10pt,pra]{revtex4}

\usepackage{amsbsy}
\usepackage{amsmath}
\usepackage{amsfonts}
\usepackage{amsthm}
\usepackage{latexsym}
\usepackage{graphicx}
\usepackage{bm}

\usepackage{fancyhdr}

\usepackage{subfigure}
\usepackage{color}

\begin{document}

\title{Heralded single photons based on spectral multiplexing and feed-forward control}

\author{M. {Grimau Puigibert}}
\affiliation{Institute for Quantum Science and Technology, and Department of Physics \& Astronomy, University of Calgary, 2500 University Drive NW, Calgary, Alberta T2N 1N4, Canada}

\author{G. H. Aguilar}
\affiliation{Institute for Quantum Science and Technology, and Department of Physics \& Astronomy, University of Calgary, 2500 University Drive NW, Calgary, Alberta T2N 1N4, Canada}
\altaffiliation{Current Address: Instituto de F\´{i}sica, Universidade Federal do Rio de Janeiro, CP 68528, 21941-972, Rio de Janeiro, Brazil}

\author{Q. Zhou}
\affiliation{Institute for Quantum Science and Technology, and Department of Physics \& Astronomy, University of Calgary, 2500 University Drive NW, Calgary, Alberta T2N 1N4, Canada}

\author{F. Marsili}
\affiliation{Jet Propulsion Laboratory, California Institute of Technology, 4800 Oak Grove Drive, Pasadena, California 91109, USA}

\author{M. D. Shaw}
\affiliation{Jet Propulsion Laboratory, California Institute of Technology, 4800 Oak Grove Drive, Pasadena, California 91109, USA}

\author{V. B. Verma}
\affiliation{National Institute of Standards and Technology, Boulder, Colorado 80305, USA}

\author{S. W. Nam}
\affiliation{National Institute of Standards and Technology, Boulder, Colorado 80305, USA}

\author{D. Oblak}
\affiliation{Institute for Quantum Science and Technology, and Department of Physics \& Astronomy, University of Calgary, 2500 University Drive NW, Calgary, Alberta T2N 1N4, Canada}

\author{W. Tittel}
\affiliation{Institute for Quantum Science and Technology, and Department of Physics \& Astronomy, University of Calgary, 2500 University Drive NW, Calgary, Alberta T2N 1N4, Canada}

\date{\today }

\begin{abstract}
We propose and experimentally demonstrate a novel approach to a heralded single photon source based on spectral multiplexing (SMUX) and feed-forward-based spectral manipulation of photons created by means of spontaneous parametric down-conversion in a periodically-poled LiNbO$_3$ crystal. As a proof-of-principle, we show that our 3-mode SMUX increases the heralded single-photon rate compared to that of the individual modes without compromising the quality of the emitted single-photons. We project that by adding further modes, our approach can lead to a deterministic SPS.
\end{abstract} 

\maketitle



Photonic quantum information processing promises delivering optimal security for sensitive communication \cite{Gisin2007}, solving certain computational problems much faster than classical computers \cite{Shor97, Monroe2010}, and estimating physical parameters with significantly improved resolution \cite{Giovannetti06}. Many of these applications rely on sources of deterministic (on-demand) and near-perfect single photons \cite{Knill01}. 
%

The most common realization of a single-photon source (SPS) is based on the generation of correlated pairs of photons (usually coined idler and signal photons) followed by the detection of one member of the pair (henceforth assumed to be the idler), which heralds the presence of the other. In this scheme a crucial step is the pair generation process, which is achieved through spontaneous parametric down-conversion (SPDC) or spontaneous four-wave-mixing (SFWM) in a nonlinear optical medium.
The experimental simplicity and versatility of such heralded sources have earned them a role in numerous quantum information applications \cite{Pan2012}. In terms of quality, these sources can produce highly indistinguishable photons, but their main limitation lies in the spontaneous nature of the pair generation. This means that single pairs of photons are generated only with a certain probability $p_{n=1}<1$ and, moreover, there is a chance of generating multiple photon pairs, which results in multi-photon emission from the SPS (for $p_{n=1} \ll 1$ the probability for generating multiple pairs is $p_{n \geq 2} \approx p_{n=1}^2$) and hence non-pure single photon states. If the collection and detection efficiency of the idler photon is very large, one can use photon-number resolving detectors \cite{Lita08,Allman15} to increase the purity, but the maximum single-photon emission probability will still be limited to $p_{n=1}=0.25$ (assuming a thermal distribution)\cite{Christ12}.

As an alternative to heralded SPSs (HSPSs) based on photon pairs, efforts have been directed to implement SPSs using single emitters. Such conceptually  simple sources are capable of deterministically emitting a single photon within the emitter's excitation lifetime, and have been demonstrated in different physical systems such as diamond colour centers \cite{Babinec10}, single molecules \cite{Lounis00} and quantum dots \cite{Michler00, Shields07}. This approach has led to nearly deterministic generation of light with nearly perfect single-photon character. The main challenge with single-emitter sources has been the lack of indistinguishability between photons emitted by different sources, or even by the same source at different times. Though progress in fabrication methods and active-control of the properties of emitters may solve this problem, it currently poses a significant drawback for practical applications.  

A promising avenue to overcome the limitations of both of the above sources is based on revisiting the heralded SPSs. The scheme proposed in \cite{Migdall02, Shapiro07} realizes in principle a deterministic SPS by actively multiplexing many non-deterministic heralded photon sources that emit photons in different modes.
In this scheme, the detection of an idler photon in any mode out of the chosen set heralds the presence of a signal photon in a corresponding optical mode. Then, using a feed-forward signal from the idler photon detector, the signal mode is mapped onto a (predetermined) single mode. 
In doing so, the probability to generate a single pair in at least one of the modes increases linearly with the number of modes $m$ (for small $p_{n=1}$), and the photon emission probability after feed-forward-based mode mapping of the signal mode can hence be made to approach unity. At the same time, the probability for multi-pair emissions into the feed-forward-mapped output mode also increases proportionally to $m$, which means that the ratio $p_{n=1}/p_{n=2}$ after mode-mapping, and hence the single-photon purity, remains constant. Though, initially, this seems unsatisfactory, it turns out to be far superior to the scaling for an individual heralded SPS. In this case, if the same increase of $p_{n=1}$ by a factor of $k=m$ were to be achieved by increasing the SPDC pump-power, it would result in an increase of $p_{n \geq 2}$ by a factor of $k^2$. Hence the ratio of $p_{n=1}/p_{n=2} \propto 1/k$. In other words the non-multiplexed HSPS would produced more and more multi-photons as the emission rate is increased. 

Multiplexed heralded sources have thus far been realized using spatial \cite{Ma2011,Collins2013}, temporal \cite{Xiong2016} and spatio-temporal modes \cite{Mendoza2016}. In some cases, they have shown to outperform non multiplexed sources in terms of throughput and quality. However, scaling up the number of  modes in the employed degrees of freedom requires more resources, and generally impacts the overall performance. In the case of spatial multiplexing, each additional mode requires an independent source and an added switching connection that induces some amount of extra loss \cite{Bonneau15}. Temporal multiplexing does not necessarily consume more physical resources but does inevitably encroach on the repetition rate of the source and hence limits the single photon throughput. 

Here, we propose and demonstrate an SPS based on a novel spectral multiplexing (SMUX) scheme in which the source requirements and the system loss are independent of the number of modes being multiplexed.
The scheme is based on defining spectral modes within the broadband spectrum of an SPDC pair source and applying a feed-forward frequency-shift operation on the heralded photon. We experimentally show that the single-photon character is preserved by measuring a heralded auto-correlation function $g^{(2)}_{\mathrm{H},0} \ll 1$ for the heralded photons with and without multiplexing and feed-forward control. Moreover, directly comparing the multiplexed and non-multiplexed output we deduce that, as expected, the heralded single-photon emissions increase linearly with the number of modes. This allows compensating for the additional loss caused by the non-ideal elements used for the feed-forward operation for as few as three modes. 

\begin{figure*}
\centering
\includegraphics[width=\textwidth]{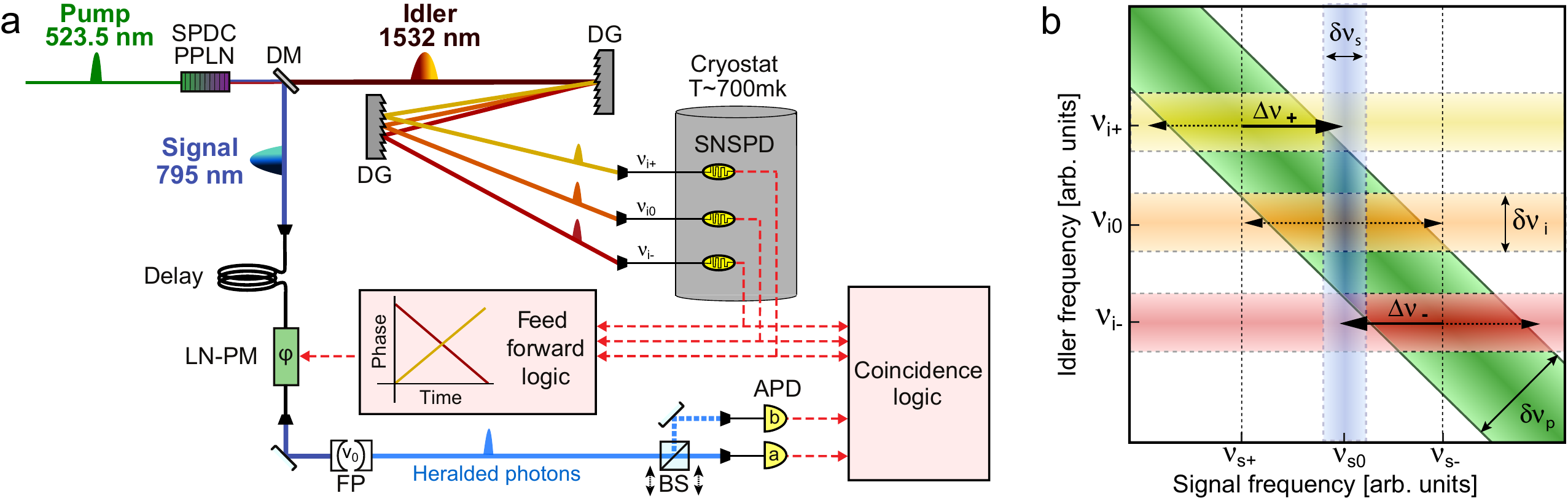}
\caption{\textbf{a. Schematic.} Pairs of of broadband (350~GHz) photons, at 1532~mm and 795~nm, are created by SPDC and separated using a dichroic mirror (DM). Three modes are defined by diffraction of the idler photon using a pair of orthogonally oriented $50\times 50$~mm square diffraction gratings (DGs) with 600 lines/mm. After detection on a superconducting nano-wire single-photon detector (SNSPD) cooled to 700~mK in a closed-cycle cryostat, the information about the frequency of the idler photon is processed in the feed-forward logic unit and a ramp generator creates an electric signal that is sent to the lithium niobate phase modulator (LN-PM). By applying different linear phase ramps, the heralded signal photons are frequency shifted and transmitted through the FP cavity. They are subsequently detected by a silicon APD-a, and a coincidence unit establishes correlations with detections in the SNSPDs. A removable 50/50 BS in the 795~nm path along with a second APD-b allows for measurement of the autocorrelation of the signal field. 
\textbf{b. Joint spectral amplitude and concept.} In green, the joint spectral amplitude for the photons produced by SPDC. The vertical and horizontal  dashed lines delimit the spectral width of the filters for the signal and idler photons, respectively. The arrows show, pictorially, how the frequency shifting acts on the spectrum of the signal photons.}
\label{fig:setup}
\end{figure*}

The experimental implementation of our scheme is shown in Fig.~\ref{fig:setup}a. A pulsed laser creating 18~ps-long pulses centred at 523.5~nm wavelength pumps a 2~cm long periodically-poled lithium-niobate (PPLN) crystal to produce 350~GHz wide, frequency non-degenerate photon pairs composed of signal photon centred at 795~nm and the idler at 1532~nm.
The spectral distribution of the pairs is conveniently illustrated by their joint spectral amplitude (JSA) shown in Fig.~\ref{fig:setup}b. The JSA represents the probability amplitude to detect a pair of photons with photons of frequencies $\nu_{s}$ and $\nu_{i}$. Since each pair of photons has to satisfy energy conservation, the JSA is confined to the diagonal band (green region), whose cross-sectional width is given by the spectral width of the pump laser $\delta \nu_p$. (The phase-matching condition is assumed to be less restrictive than the energy conservation, and is not considered in this pictorial representation. This assumption reflects standard experimental conditions.) 

The signal photons are transmitted to an optical delay line while the idler photons are sent to a pair of diffraction gratings (DGs) that map photons with spectra centred at $\nu_{i0}=195.612~\mathrm{THz}$ (1532.59~nm), $\nu_{i+}=\nu_{i0}+19~\mathrm{GHz}$ (1532.44~nm) and $\nu_{i-}=\nu_{i0}-22~\mathrm{GHz}$ (1532.76~nm), each featuring a spectral width of 12~GHz (see Supplemental Material), onto distinct spatial modes. In the JSA of Fig.~\ref{fig:setup}b these idler spectral modes are highlighted as orange (middle), yellow (upper) and red (lower) horizontal bands. The idler photons in each mode are detected using WSi superconducting nanowire single-photon detectors (SNSPDs). When an idler photon is detected in mode $\nu_{i+}$, $\nu_{i0}$ or $\nu_{i-}$, it heralds the presence of a signal photon with a central frequency of $\nu_{s0}=377.059~\mathrm{THz}$ (795.08~nm), $\nu_{s+}=\nu_{s0}-19~\mathrm{GHz}$ (795.12~nm) or $\nu_{s-}=\nu_{s0}+22~\mathrm{GHz}$ (795.03~nm), respectively.

The heralding signals from the output of the SNSPDs are processed by a logic circuit that triggers the creation of a feed-forward signal in the form of a 700~ps long pulse with suitable, linearly changing voltage. This ramp signal is applied to the electrical input of a lithium-niobate (LiNbO$_3$) phase-modulator (LN-PM) that is optically connected to the output of the 512~ns-duration delay line for the signal-photon. As we will demonstrate below, the LN-PM actively shifts the spectrum of the heralded signal photon to a spectral band --- shown as vertical blue area in the JSA --- determined by the transmission of a Fabry Perot cavity (FP) with bandwidth $\delta \nu_{s} = 6$~GHz. 
Note that the frequency shifts are applied to the entire spectrum of the signal photons, as indicated by the dotted arrows in Fig.~\ref{fig:setup}b. As a consequence, signal photons not corresponding to the heralded spectral band are shifted out of the cavity resonance and thus rejected.
Finally, the signal photons are detected by a silicon avalanche photodiode (APD) and a logic circuit records  coincidences with the heralding signals from all SNSPDs. 


%

To show how the frequency-shifting is realized by means of the LN-PM, we consider a pulse of light described in the slowly-varying-envelope approximation by $\mathcal{E}_\mathrm{in}(x,t)= |\mathcal{E}(x,t)|\exp\left(i2\pi\nu_{s0} t -ikx \right)$.
The electro-optic effect in the LN-PM translates a linearly changing voltage signal $V(t)=\mathcal{A} t $ --- applied during the passage of the optical pulse --- into a linear phase-ramp $\varphi(t)=\pi V(t)/V_{\pi} = \mathcal{A} t \pi/V_{\pi}$ of the optical pulse. Here, $\mathcal{A}$ is the slope of the voltage-signal and $V_{\pi}$ is the $\pi$-voltage of the LN-PM i.e. the voltage required to obtain a $\pi$ phase-shift. 
Hence, upon exiting the LN-PM, the pulse is described by  $\mathcal{E}_\mathrm{out}(x,t)= |\mathcal{E}(x,t)| \exp\left(i[2\pi\nu_{s0} t+ \varphi(t)]-ikx\right) =  |\mathcal{E}(x,t)| \exp\left(i2\pi[\nu_{s0} + \mathcal{A}/(2V_{\pi})] t -ikx\right) $\cite{Brecht15}. This highlights that the output pulse is frequency shifted by $\Delta \nu = \mathcal{A}/(2 V_{\pi})$ --- which is the ratio of the voltage slope to the $2 \pi$-voltage --- while the temporal shape is unchanged. 
In our experiment a heralding signal generated by the detection of an idler photon with frequency $\nu_{i+}$, $\nu_{i-}$, or $\nu_{i0}$ triggers the application of either a negative ramp with $\mathcal{A}_{-}\approx -70$~V/ns, a postive ramp with $\mathcal{A}_{+}\approx 53$~V/ns, or no ramp ($\mathcal{A}_{0}=0$~V/ns), respectively (see Supplemental Material for details).

%

\begin{figure}
\centering
\includegraphics[width=0.9\columnwidth]{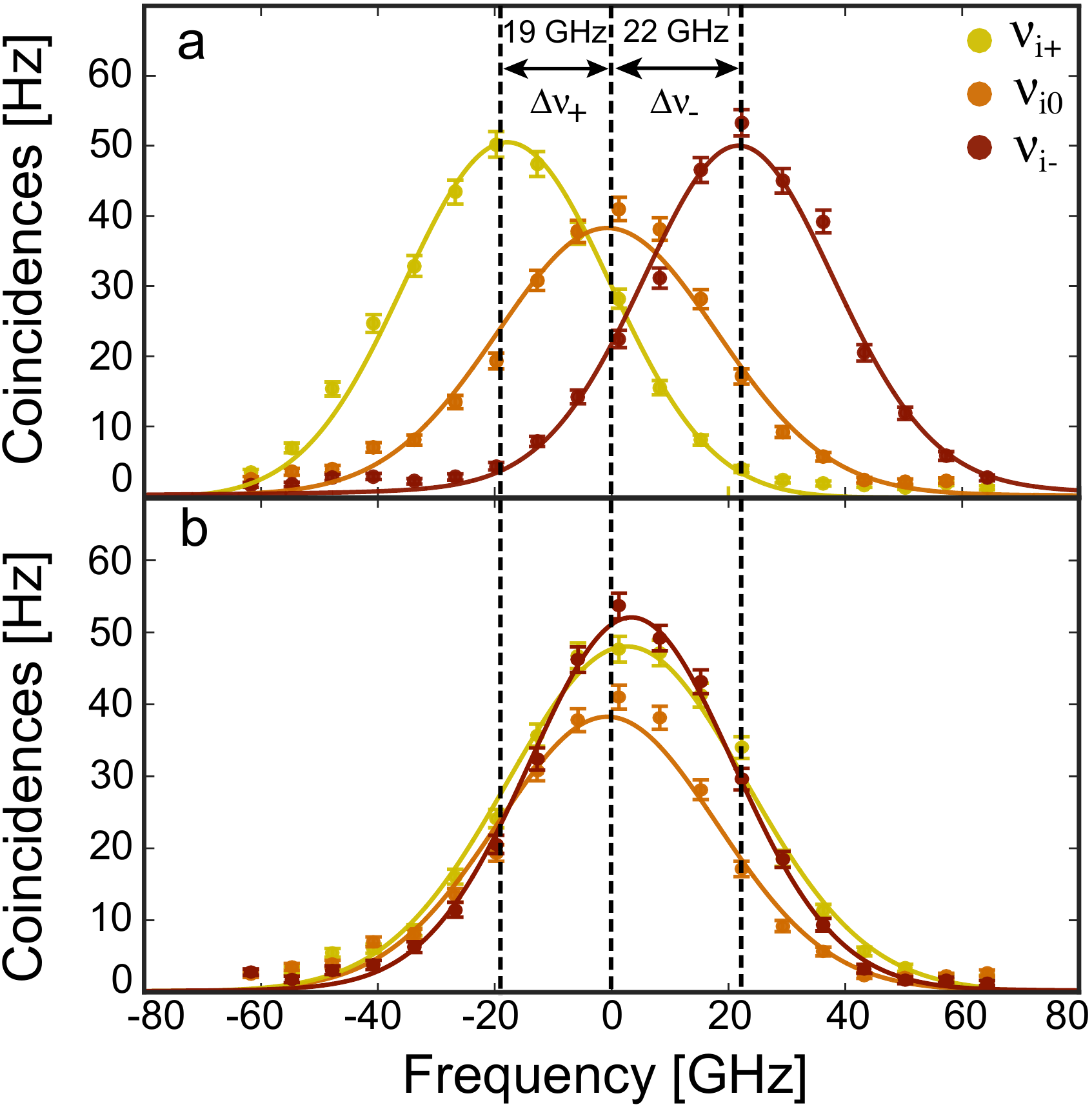}
\caption{Coincidences count rates without (\textbf{a}) and with (\textbf{b})  frequency shifting. 
The horizontal axis correspond to the relative frequency difference of the cavity resonance with respect to $\nu_{s0}$.
Coincidence rates between the signal and (heralding) idler photons at  frequencies $\nu_{+i}$, $\nu_{0i}$ and $\nu_{-i}$ are presented in yellow, orange and red, respectively. The solid lines are fits using $\mathcal{F} =  \mathcal{G}_{\nu_{s}}\star\mathcal{H}_{\nu_{i}}\star\mathcal{I}_{\nu_{p}}$, where $\mathcal{G}_{\nu_{s}}$ is a Lorentzian  function describing the frequency response of the FP cavity; $\mathcal{H}_{\nu_{i}}$ and  $\mathcal{I}_{\nu_{p}}$  are Gaussians characterizing the filtering by the DGs and the spectrum of the pump laser, respectively; and $\star$ denotes convolution.} 
\label{fig:shifting}
\end{figure}

We first characterize the spectral difference between signal photons depending on which idler photon serves as a herald. Towards this end we measure the heralded single photon rates for each idler frequency mode ($\nu_{i+}$, $\nu_{i0}$ and $\nu_{i-}$) while tuning the resonance frequency of the FP cavity that acts on the signal photon. To centre the spectral transmission of the FP at $\nu_{s0}$, we maximize the coincidence counts when heralding exclusively with idler photons detected in mode $\nu_{i0}$. Note that although the LN-PM is part of the measurement set-up, it is not active. The  results are shown in Fig. \ref{fig:shifting}a. For all modes, we measure bandwidths of the heralded photon spectra of around $37$ GHz, which matches the convolution of the pump laser bandwidth $\delta \nu_{p}=24$~GHz with the filter bandwidths $\delta \nu_{s}=6$~GHz and $\delta\nu_{i}=12$~GHz. Furthermore, we find that the maximum coincidence rate for each pair of photons is at the relative frequency differences of $\Delta \nu_{-}= -19$~GHz and $\Delta \nu_{+} = +22$~GHz. 

Next, we assess the performance of the linear ramp frequency shifting (LRFS) by again recording the heralded-photon spectra for each idler frequency-mode, but now with the corresponding feed-forward signal applied to the LN-PM. For these measurements, the resonance frequency of the FP cavity remains fixed at $\nu_{s0}$. As expected, the spectra shown in Fig.~\ref{fig:shifting}b now completely overlap. Moreover, by comparing with the results in Fig.  \ref{fig:shifting}a, we observe that the detection rates with and without LRFS are essentially equal. This shows that our setup is capable of applying the on-demand frequency shift at the single-photon level with nearly $100\%$ efficiency.

\begin{figure}
\centering
\includegraphics[width=0.95\columnwidth]{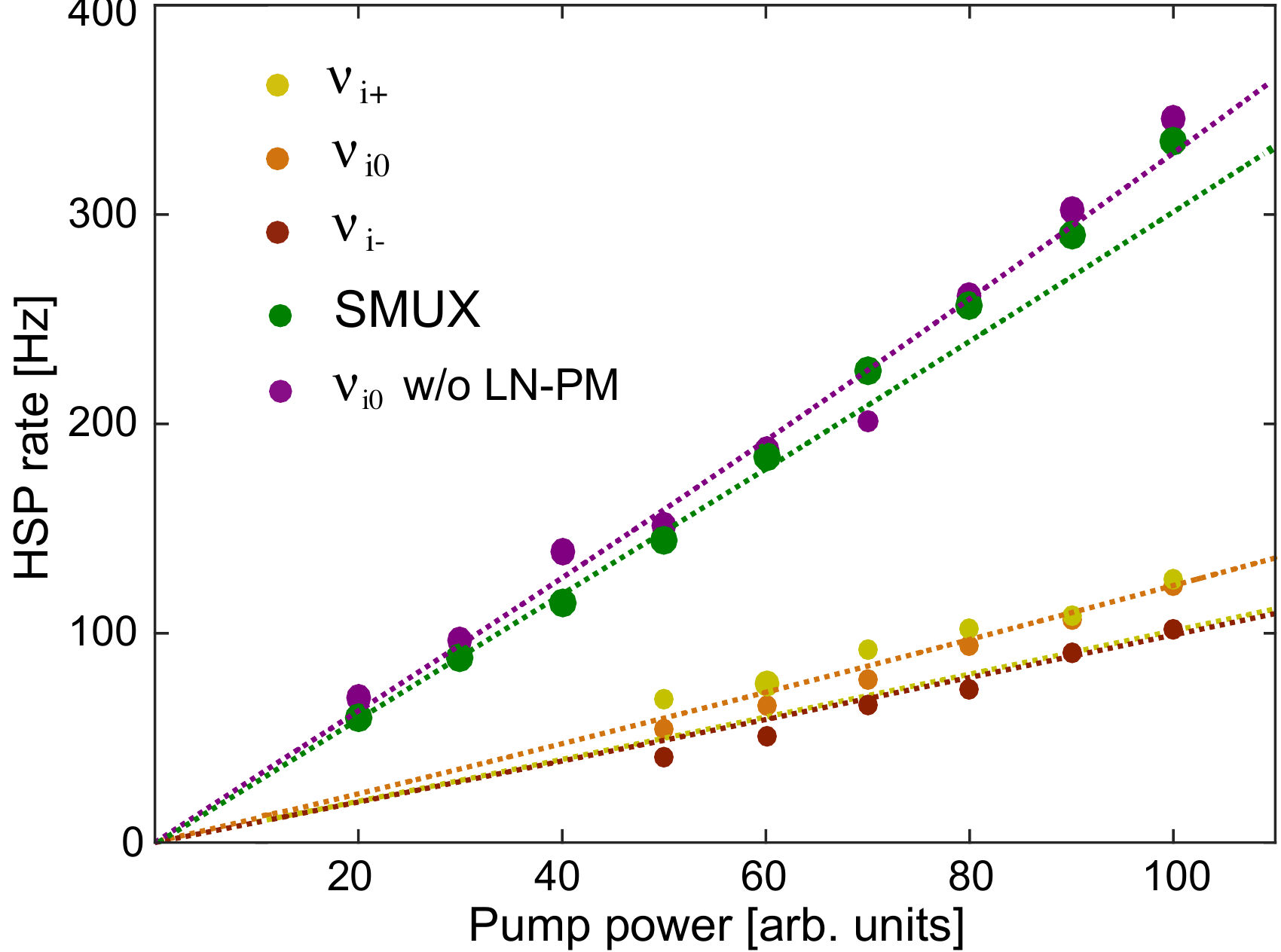}
\caption{\textbf{a} HSP rate versus pump power for the individual frequency modes (red, yellow and orange circles), for SMUX of three modes (green circles) and for the $\nu_{0}$ mode without the LN-PM in the signal photon mode (purple circles). Lines are linear fits to the data. 
} 
\label{fig:figureCAR}
\end{figure}

Now we activate the full setup and evaluate the performance of our frequency multiplexed heralded source in view of the requirements of an ideal single photon source. First, we measure the heralded single photon (HSP) rate as a function of the pump power with and without multiplexing. The red, yellow and orange circles in Fig. \ref{fig:figureCAR} show the rates of signal photons (after spectral filtering by the FP cavity) heralded by the individual frequency modes at $\nu_{i+}$, $\nu_{i0}$ and $\nu_{i-}$. The multiplexed HSP rate for the SMUX source (green circles) is about 2.7 times larger than that of the average of the individual sources, i.e. a significant improvement over that for the individual modes. Yet, to make a fair comparison of the effect of the SMUX, we compare the multiplexed HSP rate to the rate when heralding with only the $\nu_{i0}$ mode and with the LN-PM removed, which, for the specific modulator used in our experiment, adds 5 dB loss and is not needed in the case of using only a standard HSPS without multiplexing. However, we keep the spectral filtering elements on both idler and signal fields as these are necessary to achieve pure states \cite{Mosley08}. The rate obtained with the thus modified source (purple circles) is similar to the SMUX rate, which can be explained by the increased transmission due to having removed the LN-PM compensating for the lack of multiplexing. Hence, in our experiment, which employs 3 modes, spectral multiplexing does not yet create an advantage in view of the HSP rate and in creating a deterministic source. However, we emphasize that increasing the number of modes would neither increase the system loss nor require additional elements in the signal mode. Hence, we can assume that the HSP rate will continue to increase as more spectral modes are added, rapidly surpassing that of the non-multiplexed source. 

Next, we verify that multiplexing maintains the single photon character of our light source. One of the most common methods for this is to determine its purity \cite{Somaschi16}, which, for a heralded source, is generally quantified in terms of its heralded auto-correlation function. To measure this, we direct the heralded signal photons after the FP cavity through a 50/50 beam splitter (BS), and record the individual counts $C_\mathrm{H}^\mathrm{(i)} (i \in \{a,b\})$, in as well as coincidences $C_\mathrm{H}^\mathrm{(ab)}$ between, the detectors placed at its two outputs (see Fig.~\ref{fig:setup}a). Together with the total number of heralding signals, $H$, we define the heralded auto-correlation as $g^{(2)}_{H}(0)=C_\mathrm{H}^\mathrm{(ab)}H/(C_\mathrm{H}^\mathrm{(a)}C_\mathrm{H}^\mathrm{(b)})$ (see Supplemental Material).
Since detection in both detectors  can only occur if two or more photons are present we expect $C_\mathrm{H}^\mathrm{(ab)}/H \approx p_{n=2}^\mathrm{H}/2$, where $p_{n=2}^\mathrm{H}$ is the heralded two-photon probability. On the other hand, the individual detector counts are dominated by single photon events such that e.g $C_\mathrm{H}^\mathrm{(a)}/H \approx p_{n=1}^\mathrm{H}/2$ (assuming $p_{n=1} \gg p_{n=2}$). Clearly these considerations hold for the multiplexed and non-multiplexed sources alike (see Supplemental Material). Hence, the autocorrelation function is approximated by $g^{(2)}_{H}(0) \approx 2 p_{n=2}^\mathrm{H}/(p_{n=1}^\mathrm{H})^2$, which provides a direct relation to the photon-statistics. In particular, $g^{(2)}_{H}(0) \sim 0$ would indicate 
$p_{n=2}^\mathrm{H}\sim 0$ and $p_{n=1}^\mathrm{H} > 0$.

Operating the source at maximum power with the LN-PM (and hence the multiplexing) removed we obtain a value of $g^{(2)}_{H}(0)=0.05 \pm 0.01$. With the SMUX activated and using the same pump power, we measure $g^{(2)}_{H}(0)= 0.06 \pm 0.01$, which equals the value without multiplexing within experimental uncertainty. If the LN-PM loss were lower we would thus achieve a larger HSP rate without increasing the $g^{(2)}_{H}(0)$ (see Supplemental Material). Both experimentally measured $g^{(2)}_{H}(0) \ll 1$, which indicates a highly pure single photon character of the output mode. Hence, we establish that the SMUX maintains the single photon nature of the source. This conclusion is supported by additional measurements of the so-called coincidence to accidental ratio (CAR) measured with and without the SMUX \cite{Zhou17}.

We did not directly measure the degree of distinguishability of single photons from our source, but based on the experimental setup, we conjecture they must be nearly indistinguishable. The reason for this is that the photons emitted by our SMUX source pass through a FP-cavity and are coupled into single mode fibre, which removes any distinguishability in the frequency and spatial degrees of freedom. Finally, the fact that all signal photons travel along the same path, from the point where they are created to the point where they are detected, means that their arrival times and polarization are the same (irrespective of which spectral mode they belong to), provided that chromatic dispersion and polarization mode dispersion can be ignored.

In conclusion, we have generalized the idea of multiplexing supplemented by feed-forward control with the goal of creating a high-quality heralded SPS from the spatial and temporal degrees-of-freedom to the spectral domain. We have demonstrated that multiplexing of three spectral modes leads to the expected increase in the heralded single photon rate while keeping the purity -- the noise caused by multiple-pair emission -- constant.
To achieve these results we have implemented on-demand frequency shifting of single photons over approximately $\pm20$~GHz with nearly 100$\%$ efficiency by driving a commercially available LiNbO$_3$ phase modulator with a linear voltage ramp. Unlike in the case of temporal or spatial multiplexing \cite{Bonneau15}, the required resources  -- with the exception of the single-photon detectors -- do not depend on the number of modes, thus making this approach appealing for further development towards an ideal SPS.

There is a number of avenues to increase the heralded single-photon rate of our source, the simplest being to employ a phase-modulator with reduced loss. For instance, the use of a readily available modulator with 1.5 dB loss -- 3.5 dB less than in our current modulator ---  would result in an improvement of the rate by about a factor of 2.5. At a more fundamental level, more spectral modes must be multiplexed. This can be achieved both by increasing the spectral mode density (e.g. by taking advantage of commercially available ultra dense wavelength division multiplexers with 6 GHz bandwidth), as well as by improving the total bandwidth -- the latter necessitating improvement of the maximum frequency shifs using higher-bandwidth electronics \cite{Wright16,Li16,Fan16,Albrecht14}. Assuming 6 GHz wide spectral channels and an 8-fold increase of the shifting range to 350~GHz (the spectral width of the photons created by our source), it seems feasible to reach 60 spectral channels. This will suffice for creating at least one heralding event per pump pulse and for making the source as close to deterministic as loss in the signal path allows \cite{Christ12}. And as the SPDC spectrum is determined by the length of the non-linear crystal and phase-matching conditions, even larger bandwidths and thus more modes may be feasible.

To finish this paper, let us note that spectral multiplexing has also been proposed to improve the rate of entanglement distribution in a quantum repeater architecture \cite{Sinclair14}. However, the proof-of-principle demonstration in \cite{Sinclair14} was still lacking the feed-forward control. Our current demonstration therefore also establishes an important element in this architecture.

\acknowledgements
The authors thank Vladimir Kiselyov for technical support and Raju Valivarthi for discussions. This work was funded through Alberta Innovates Technology Futures (AITF), and the National Science and Engineering Research Council of Canada (NSERC). VBV and SWN acknowledge partial funding for detector development from the Defense Advanced Research Projects Agency (DARPA) Information in a Photon (InPho) program. Part of the detector research was carried out at the Jet Propulsion Laboratory, California Institute of Technology, under a contract with the National Aeronautics and Space Administration. WT furthermore acknowledges funding as a Senior Fellow of the Canadian Institute for Advanced Research (CIFAR).



\end{document}


\title[a]{Heralded single photons based on spectral multiplexing and feed-forward control\\ Supplemental Material}%

\author{M. {Grimau Puigibert}}
\affiliation{Institute for Quantum Science and Technology, and Department of Physics \& Astronomy, University of Calgary, 2500 University Drive NW, Calgary, Alberta T2N 1N4, Canada}

\author{G. H. Aguilar}
\affiliation{Institute for Quantum Science and Technology, and Department of Physics \& Astronomy, University of Calgary, 2500 University Drive NW, Calgary, Alberta T2N 1N4, Canada}
\altaffiliation{Current Address: Instituto de F\´{i}sica, Universidade Federal do Rio de Janeiro, CP 68528, 21941-972, Rio de Janeiro, Brazil}

\author{Q. Zhou}
\affiliation{Institute for Quantum Science and Technology, and Department of Physics \& Astronomy, University of Calgary, 2500 University Drive NW, Calgary, Alberta T2N 1N4, Canada}

\author{F. Marsili}
\affiliation{Jet Propulsion Laboratory, California Institute of Technology, 4800 Oak Grove Drive, Pasadena, California 91109, USA}

\author{M. D. Shaw}
\affiliation{Jet Propulsion Laboratory, California Institute of Technology, 4800 Oak Grove Drive, Pasadena, California 91109, USA}

\author{V. B. Verma}
\affiliation{National Institute of Standards and Technology, Boulder, Colorado 80305, USA}

\author{S. W. Nam}
\affiliation{National Institute of Standards and Technology, Boulder, Colorado 80305, USA}

\author{D. Oblak}
\affiliation{Institute for Quantum Science and Technology, and Department of Physics \& Astronomy, University of Calgary, 2500 University Drive NW, Calgary, Alberta T2N 1N4, Canada}

\author{W. Tittel}
\affiliation{Institute for Quantum Science and Technology, and Department of Physics \& Astronomy, University of Calgary, 2500 University Drive NW, Calgary, Alberta T2N 1N4, Canada}

\date{\today}
\maketitle

\section*{Spectral mode selection}

In order to select the different spectral modes for the heralding photons, i.e. the idler modes at $\sim 1532$~nm, we use a home-made spectrometer that is composed of two diffraction gratings (DGs) (Thorlabs GR50-0616) orthogonally oriented with respect to each other. The idler mode is expanded to a spot-size of  $\sim23$~mm in order to illuminate a large number of grooves on both diffraction gratings (recall that the spectral resolution scales linearly with the number of groves that the optical mode covers \cite{Horiba16}). After the two DGs, the different spectral modes are coupled into separate single-mode fibres. The spectral bandwidth of each mode is determined by the angular spread caused by the gratings together with the size of the spatial mode supported by the optical fibre.

\begin{figure}[h!]
   \centering
   \includegraphics[width=0.48\textwidth]{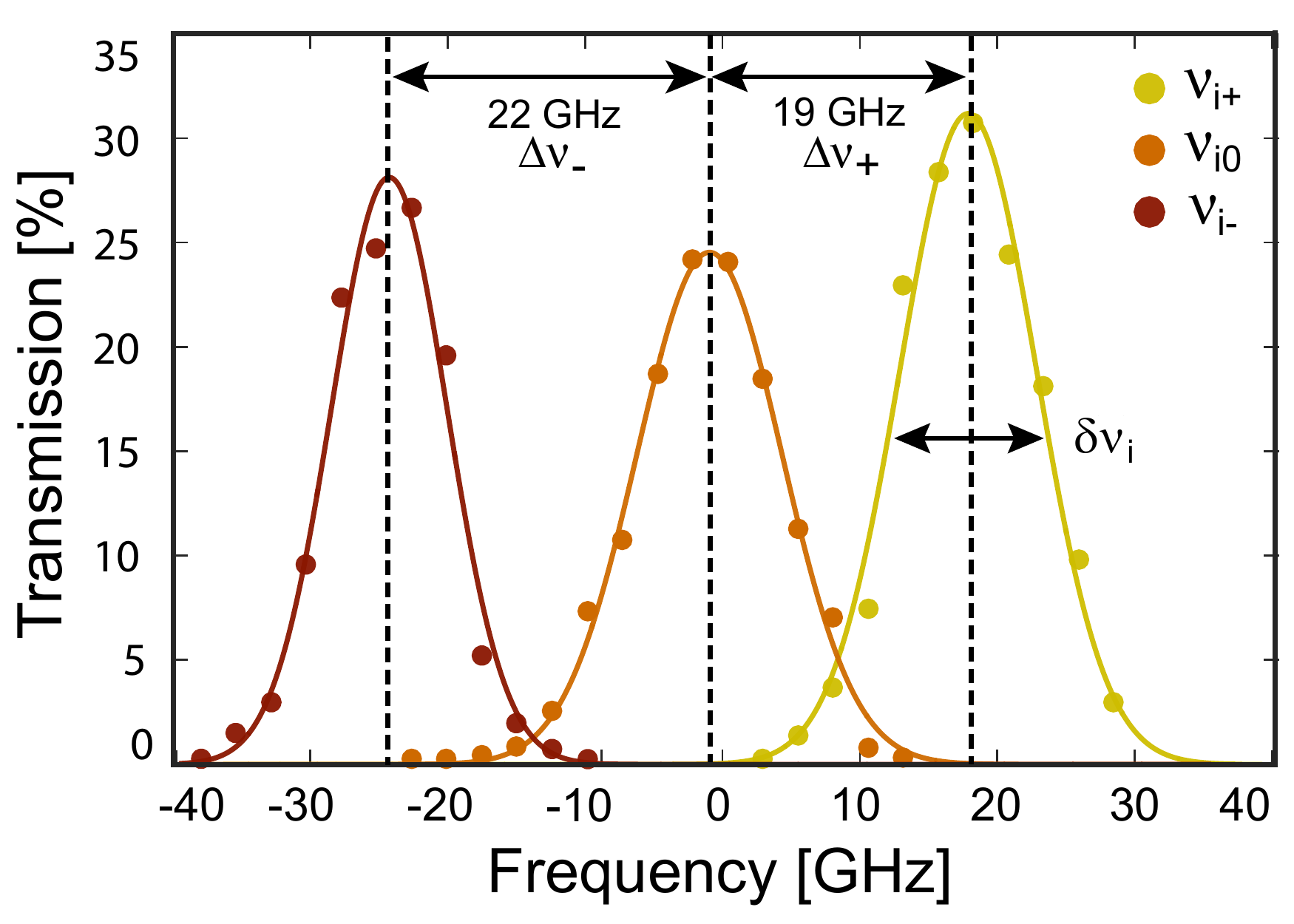}
   \caption{Spectra of the different spectral modes defined in the idler photon by the DGs. The total transmission through the spectrometer (in percentage) shown for each relative frequency mode, measured with respect the central mode $\nu_{i0}$ (1532.59~nm). The modes $\nu_{i-}$ and $\nu_{i+}$ are separated by $\Delta\nu_{+}\approx 19$~GHz and $\Delta\nu_{-}\approx 22$~GHz, respectively. The  bandwidth is $\delta \nu_i \approx 12$~GHz for the three modes. Experimental points are fitted with Gaussian functions.}    
\label{fig:spectrum}
\end{figure}


The bandwidth and transmission of each spectral mode ($\nu_{i-}$, $\nu_{i0}$ and $\nu_{i+}$) is measured by scanning the  wavelength of a tuneable continuous wave laser and recording the transmitted light intensity coupled into each of the three fibres. The results are plotted in Fig. \ref{fig:spectrum}. The bandwidth $\delta \nu_i$ is around 12 GHz for each of the three modes, and the transmissions are 25$\%$, 22$\%$ and 30$\%$ for the modes centered at $\nu_{i-}$, $\nu_{i0}$ and $\nu_{i+}$, respectively. Note that the mode overlap is negligible. The frequency separation between the modes is matched to the frequency shifts of our feed-forward system, which we describe in the next section.

\section*{Frequency-shifting of single photons and ramp-signal generation}

Frequency-multiplexing of heralded single photons requires photonic devices that actively change the frequency of the photons on-demand. As discussed in the main text, we shift the frequency of the signal photons by applying a linearly varying phase ramp by means of a lithium niobate phase-modulator (LN-PM) \cite{Laura2016}. The magnitude $\Delta\nu$ of the frequency-shift in hertz is proportional to the rate $d\phi(t)/dt$ of increase/decrease of the phase during the passage of the photon. Assuming a linearly varying voltage $V(t)=\kappa\, t$ applied to the phase-modulator, we can write
\begin{equation}
\Delta\nu = \frac{d\phi(t)}{2\pi dt}=\frac{\pi}{V_{\pi}} \frac{dV(t)}{2\pi dt} =  \frac{\kappa}{2V_{\pi}} ,
\label{eq:fs}
\end{equation}
where $V_{\pi}$ is the $\pi$-voltage of the LN-PM, i.e. the DC-voltage required to change the phase of the passing light by $\pi$ radians.~We assume that $V_{\pi}$ is independent of the rate of change of the applied voltage modulation or, in other words, that the bandwidth of the phase-modulator is greater than the relevant spectral components of the input voltage signal. Furthermore, we assume that $V_{\pi}$ is constant over the total bandwidth of the signal photons.

\begin{figure}[h!]
   \centering
   \includegraphics[width=0.48\textwidth]{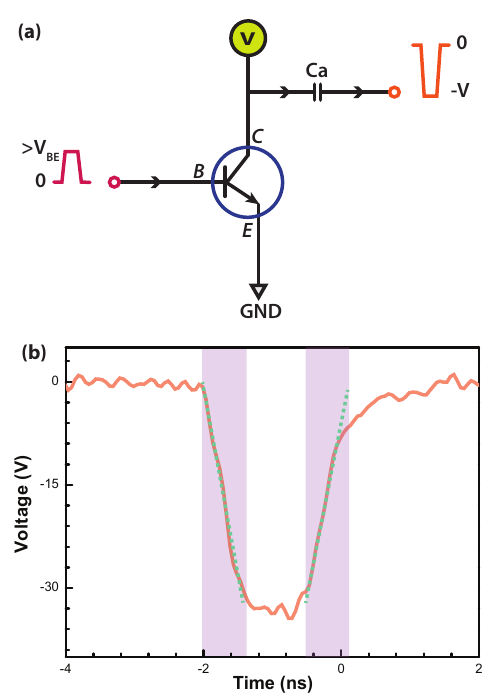}
   \caption{(a) Schematic circuit for ramp-signal generation; (b) Typical output of the generator.}
   \label{fig:rg}
\end{figure}

In our set-up the voltage signal is generated from a home-built ramp-signal generator. Figure~\ref{fig:rg}a depicts the schematic of our electronic circuit.~The key component in the circuit is a high-voltage and large-bandwidth radio frequency (RF) transistor that acts as an inverting electrical amplifier. The ramp-signal generator takes an input signal, such as an appropriately shaped single-photon detector output, and outputs a $\pi$-phase shifted (negative) high-voltage pulse with linearly increasing falling and rising edges as shown in Fig.~\ref{fig:rg}b. The pulse amplitude is -35~V, and the falling and rising edges (shaded regions) are around 700~ps long. When the temporal wave-packet of the photon overlaps with the falling or rising edge of the pulse, Eq.~(\ref{eq:fs}) predicts that the spectrum of the photon is shifted in the negative or positive direction, respectively. Since the slope of the rising and falling edges are slightly different, our set-up provides frequency shifts of $\Delta\nu_{-} = -22$~GHz and $\Delta\nu_{+} = +19$~GHz, which agree with the spectral mode filtering shown in~Fig.~\ref{fig:spectrum}.

The feed-forward control circuitry, which selects the correct value for the frequency shift depending on which idler mode provides the heralding detection, also adjusts the timing of the input pulse of the ramp generator. More specifically, the heralding signals from the $\nu_{i-}$ and $\nu_{i+}$ modes are delayed with respect to each other such that for the former, the falling (front) and for the latter the rising (back) edge of the output pulse coincides with the passage of the signal photon through the LN-PM. If a herald originates from the $\nu_{i0}$ mode no frequency shifting is required and the heralding signal is just not input into the ramp-generator.

Frequency shifting at the single-photon level has been employed in previous experimental demonstrations \cite{Laura2016, Tang2016}. To put our ramp-signal-based frequency-shifting method into context with these single-photon frequency shifters, we define a `time-bandwidth-like' product that can serve as a figure-of-merit to gauge the performance of different approaches. The `time' factor in our figure-of-merit is the maximum temporal duration $\Delta\tau$ of the optical pulse that will experience a constant frequency-shift. For a Fourier-transform-limited optical wave-packet, a larger $\Delta\tau$ indicates that a spectral mode with smaller spectral width can be defined, thus more distinct spectral modes can be obtained given the total spectral range of the frequency shifter. The `bandwidth' factor is the maximum absolute frequency-shift $\Delta \nu_{max}$ of the optical spectrum. A larger $\Delta \nu_{max}$ means a broader spectral range that a frequency shifter can cover.~Putting both of them together, a frequency shifter with larger `time-bandwidth-like' product indicates that it can manipulated more spectral modes, which is desired for spectrally multiplexed heralded single photon sources and for light-matter interfaces based, spectrally multiplexed quantum repeaters \cite{Neil2014, Erhan2016}. 

We are able to achieve a time-bandwidth product of $0.7~\mathrm{ns} \times 22~\mathrm{GHz} = 15.4$. Although larger frequency shifts have been demonstrated, the above-defined figure-of-merit in those experiments was one order of magnitude smaller.~For example the time-bandwidth product was $6.25~\mathrm{ps} \times 200~\mathrm{GHz} = 1.25$ in \cite{Laura2016} and $4~\mathrm{ps} \times 150~\mathrm{GHz} = 0.6$ in \cite{Tang2016}. The key factor that allows us to achieve the large `time-bandwidth-like' product is that the frequency of our single photons is shifted with a single ramp-signal pulse of $<2$~ns duration. As a consequence, even though we use a high-voltage signal at the LN-PM input, the root-mean-square (RMS) power applied to the LN-PM is very small. This is in contrast to the electronic signals with large RMS power that were used in \cite{Laura2016, Tang2016}.

\section*{Heralded auto-correlation function}

In this section we provide some basic insight into how the auto-correlation function can be used to gauge the performance of a heralded single photon source and how it is affected by multiplexing. The heralded auto-correlation function, $g^{(2)}_{H}(0)$, is commonly used to determine the single photon purity \cite{Somaschi16,Sekatski12} of the signal mode after heralding by the detection of  photon in the idler mode. Values of $g^{(2)}_{H}(0)$ below 1 indicate a non-classical light field with a value of 0  corresponding to an ideal single photon.

To measure the auto-correlation function, the signal mode is split equally on a beam-splitter ,and from detectors placed at the two outputs we record the individual counts $C_\mathrm{H}^\mathrm{(i)}$ ($i \in \{a,b\}$), the coincidental counts $C_\mathrm{H}^\mathrm{(ab)}$  as well as the number of heralding signals $H$ (see Fig.~1a in the main text). This allows us to define the heralded detection probabilities, e.g. $p_\mathrm{H}^\mathrm{(ab)}= C_\mathrm{H}^\mathrm{(ab)}/H$, and thus the heralded auto-correlation function as
\begin{align}
g^{(2)}_{H}(0)&\equiv \frac{p_\mathrm{H}^\mathrm{(ab)}}{(p_\mathrm{H}^\mathrm{(a)}p_\mathrm{H}^\mathrm{(b)})}
= \frac{C_\mathrm{H}^\mathrm{(ab)} H}{(C_\mathrm{H}^\mathrm{(a)}C_\mathrm{H}^\mathrm{(b)})} \nonumber \\[5pt] 
&\approx \frac{2p_{n=2}^\mathrm{H}}{(p_{n=1}^\mathrm{H})^2} \ .
\label{eq:autocorr}
\end{align}
Here $P_{n=1}^\mathrm{H}$ and $p_{n=2}^\mathrm{H}$ correspond to the heralded probabilities of having a single photon or two photons, respectively, in the signal field, i.e. in the output of the photon source. The definition in Eq.~(\ref{eq:autocorr}) does not distinguish between single mode and multiplexed operation and is thus valid for both cases. The approximation, which holds for $p_{n=1}\ll1$, provides an intuitive link between $g^{(2)}_{H}(0)$ and the photon number distribution, more specifically the ratio of the undesired multi-photon probabilities (which is usually dominated by the two-photon contribution) to the desired single-photon probability in the heralded output.

Experimentally, $p_{n=1}^\mathrm{H}$ ($p_{n=2}^\mathrm{H}$) can be estimated by counting the number of single (two) photon events and dividing it by the number of heralding signals $H$. For a multiplexed source with a degree of multiplexing $m$ and assuming each mode operating under the same conditions, the total number of single (two) photon events and the total number of heralding signals both increase proportional to $m$. This means that $p_{n=1}^\mathrm{H}$ ($p_{n=2}^\mathrm{H}$) remain the same as in the single mode case, and, consequently, the value of $g^{(2)}_{H}(0)$ is independent of the number of multiplexed modes $m$.

The advantage of using a multiplexed source versus a single mode source is that for a given value of $g^{(2)}_{H}(0)$ --- which is related to the quality of the individual modes of the source --- the rate of heralded photons in the multiplexed case ($C_\mathrm{H}\propto m p_{n=1}$) is $m$-times larger than for the non-multiplexed case ($C_\mathrm{H}\propto p_{n=1}$). If the multiplexed operation adds extra loss, $\eta$, to the signal channel, the multiplexed rate will be lowered to $C_\mathrm{H}\propto m (1-\eta) p_{n=1}$. In conclusion, for a given $g^{(2)}_{H}(0)$ a multiplexed heralded photon source will outperform a non multiplexed source in terms of photon output rates only if $(1-\eta)<(1/m)$. Due to extra loss introduced by the LN-PM in our current implementation, $(1-\eta)=0.32$. In comparison, using three modes, we find  $(1/m)=0.33$.

\section*{Efficiencies and transmissions}

Table~1 list the transmissions and efficiencies of the elements used in the experiment and identified in figure 1a of the main text. Furthermore, detector efficiencies are 60 $\%$ for the Si-APD, and $\approx$ 70 $\%$ for the SNSPDs (slight variations between different detectors are due to different external losses).

\begin{center}
\begin{table}[htp]
    \begin{tabular}{| c | c | c |}
   \hline
    Element & label & Transmission [$\%$]\\ \hline
    Optical Fiber loop & Delay & 81 \\ 
    Fabry-Perot Cavity & FP & 45\\
    LiNbO$_3$ Phase-Modulator & LN-PM & 32\\ \hline 
    \end{tabular}
     \label{tab:trans}
     \caption{Optical transmission for the different elements use in our demonstration.}
\end{table}
\end{center}